\documentclass[12pt,a4paper]{article}
\usepackage{jheppub}
\def\CH{{\mathcal H}}

\title{Entropy Current for the Fluid/Gravity Model of the Chiral Magnetic Effect}
\author{T. Ashok}
\affiliation{Department of Applied Mathematics and Theoretical Physics, University of Cambridge, Cambridge CB3 0WA, England}
\emailAdd{A.Thillaisundaram@damtp.cam.ac.uk}

\abstract{
We construct long wavelength asymptotically locally AdS$_5$ spacetimes with slowly varying (background) gauge fields which are solutions to the $U(1)^n$ Einstein-Maxwell-Chern-Simons system. These bulk spacetimes are dual to $(3+1)-$dimensional fluid flows with $n$ anomalous currents in the presence of external electromagnetic fields. We utilise the area form on the outer horizon to holographically compute an entropy current for the dual fluid to first order in boundary derivatives. Our resulting expression contains additional terms proportional to the vorticity and magnetic field and thus provides holographic confirmation of the entropy current calculated by Son and Surowka for hydrodynamics with triangle anomalies. We then restrict our bulk metric to describe the fluid/gravity model of the chiral magnetic effect (CME) and again holographically obtain the entropy current. As expected, our calculation replicates the result produced using standard hydrodynamic/thermodynamic arguments.

}

\begin{document}
\maketitle
\section{Introduction}

The gauge/gravity duality \cite{Maldacena1,Maldacena2} allows us to study the behaviour of strongly coupled field theories in terms of an equivalent description involving the gravitational dynamics of black holes in one higher spatial dimension. This dual perspective has proved remarkably useful in extracting non-equilibrium properties of strongly coupled field theories; here, standard field theoretic techniques have been much less successful. A now standard holographic approach is to consider linear perturbations about a static black hole spacetime which can then inform us about the small amplitude transport properties of the dual field theory \cite{Son1,Son2,Son3,Son4}. These holographic linearised hydrodynamical calculations have been widely applied with great success \cite{Gubser1,Hartnoll1}.

A natural question to ask is whether the progress achieved using AdS/CFT in the study of transport properties can be extended beyond the linear level. From the dual gravitational viewpoint, this is analogous to trying to find time-dependent solutions to the full nonlinear gravitational equations - clearly not a straightforward problem. Yet, recently, significant progress was made in \cite{B1} where the authors  obtained the bulk spacetime dual to \textit{nonlinear} fluid dynamics. Fluid dynamics \cite{L} is the long wavelength effective theory of a locally equilibriated field theory; and its governing equations, known as the Navier-Stokes equations, simply follow from conservation of the stress tensor. In \cite{B1}, the authors considered field theories in the 't Hooft limit and at infinite coupling; the holographic duals of such field theories are generically solutions of two derivative Einstein gravity coupled to other fields. To obtain the bulk dual of the simplest case of an uncharged fluid, they perturbatively constructed asymptotically AdS$_5$ tube-wise black brane solutions to Einstein gravity with negative cosmological constant to second order in a boundary derivative expansion. By considering appropriate projections of the bulk gravitational equations, they showed explicitly that the boundary field theory obeys the conformal relativistic Navier-Stokes equations. Their explicit map between the dynamics of fluids and that of black holes has become known as the fluid/gravity correspondence \cite{R2,R3}; and the results of their seminal paper \cite{B1} have since been generalised in various interesting directions \cite{B2,Haack,B3,Ashok,B4,Skenderis3,Erdmenger,Banerjee,B5,Herzog,Myers,Oz,Oz2,Liu3,Kirsch}. 

The particular extension that we are interested in here is the fluid/gravity model of the chiral magnetic effect (CME) \cite{Kirsch}. The chiral magnetic effect \cite{Warringa} is a phenomenon where a background magnetic field induces a current in the direction of the field in the presence of imbalanced chirality; such imbalances in chirality can arise in topologically nontrivial gluonic configurations in QCD due to the axial anomaly, for example. Early work on the CME derived this current for equilibrium systems \cite{Warringa}. More recently, a hydrodynamic model of the CME was constructed \cite{Sad,Pu}; these results are based on the arguments presented in \cite{Surowka} where the authors considered (holographic) hydrodynamics with triangle anomalies. The results of \cite{Surowka} represent one of the notable triumphs of AdS/CFT; and, as their work is relevant to our discussion, we now pause to elaborate on their paper and to provide some background. 

For generic hydrodynamic models, the currents can be expressed as a sum of terms allowed by symmetries, each with an attached transport coefficient. However, in canonical textbook examples \cite{L}, some of these terms were disallowed on the grounds that they were inconsistent with the second law of thermodynamics. However, most surprisingly, in recent holographic calculations involving the bulk dual of a charged fluid \cite{Erdmenger,Banerjee}, one of these coefficients was found to be nonzero, implying an incompatibility with the second law. In \cite{Surowka}, the authors resolved this issue. By considering fluid dynamics with triangle anomalies, they showed that that these problematic transport coefficients can be included if we allow the entropy current to be modified by the inclusion of additional terms; further, such additional terms are in fact \textit{required} by the presence of triangle anomalies. Their findings have challenged standard lore and have led to a modification of the canonical equations of fluid dynamics. 

In the simplest case of a charged fluid with one $U(1)$ current (and a $U(1)^3$ anomaly), the additional terms to the charge current represent contributions to the current in the direction of the vorticity and external magnetic field (if one is present). And it is the term proportional to the magnetic field that has been connected to the chiral magnetic effect in heavy ion collisions. Given that heavy ion collisions probe physics at strong coupling, it is useful to establish holographic models describing the CME; here, calculations at strong coupling can be done using fairly straightforward gravitational methods. The fluid/gravity model of the CME mentioned above \cite{Kirsch} is an example of a successful holographic model (see \cite{Hol1,Hol2,Hol3,Hol4,Hol5,Hol6,Hol7,Hol8} for earlier attempts). In \cite{Kirsch}, the authors holographically modelled a hydrodynamic system involving two currents, denoted by $U(1)_A$ and $U(1)_V$ respectively, in the presence of a background magnetic field. These currents represent the axial and vector currents in QCD, $j_5^\nu=\bar{q}\gamma^\nu\gamma^5 q$ and $j^\nu=\bar{q}\gamma^\nu q$, and we label their associated chemical potentials by $\mu$ and $\mu_5$. The chemical potential $\mu_5$ for the $U(1)_A$ axial current, which the authors modelled as anomalous, can be thought of as representing the imbalance in chirality due to the axial anomaly. The presence of the anomalous axial current requires the addition of the extra transport coefficient signifying the existence of a current in the direction of the magnetic field. Their fluid/gravity model correctly reproduced the results for the transport coefficients given in \cite{Sad,Pu} and thus provides holographic confirmation of their hydrodynamic extension of the CME.

In this paper, we extend the results of \cite{Kirsch} by holographically computing the entropy current for the hydrodynamic model of the CME. We first demonstrate that the bulk solutions presented in \cite{Kirsch} possess a regular event horizon and we determine its location; then, utilising the area form on the horizon, we construct a gravitational entropy current. Mapping this to the boundary along ingoing null geodesics allows us to determine an entropy current for the corresponding hydrodynamic model of the CME defined on the boundary. The area increase theorem of general relativity guarantees the positive divergence of this entropy current.

More concretely, we first obtain long wavelength bulk solutions dual to a fluid with \textit{multiple anomalous currents} in the presence of \textit{external electromagnetic fields} - this generalises previous work which considered more specific configurations \cite{Hur,Erdmenger,Megias,Banerjee,Kirsch}. The dynamics of such bulk spacetimes are governed by the $U(1)^n$ Einstein-Maxwell action with a Chern-Simons term:
\begin{equation}
S=\frac{1}{16\pi G_5} \int d^5x \sqrt{-g}\left[R-12-F_{MN}^a F^{aMN}+\frac{S^{abc}}{6\sqrt{-g}}\epsilon^{PKLMN}A_P^a F_{KL}^b F_{MN}^c\right]
\end{equation}
We calculate the location of the regular event horizon for such fluid bulk duals and determine the expression for the entropy current. These results are in themselves already novel generalisations of existing work - the location of the event horizon and the form of the entropy current thus far have only been calculated for bulk duals of fluids with a single charge without any background electromagnetic fields \cite{Spalinski}. Our expression for the entropy current contains additional terms proportional to the vorticity and magnetic field, and is completely consistent with the results of \cite{Surowka}. We then specify to the fluid/gravity model of the CME and holographically calculate its entropy current.

This paper is organised as follows. In sections \ref{2} and \ref{3}, we review the methodology introduced in \cite{Harvey} to determine the location of the event horizon and the expression for the entropy current for a general fluid dynamical metric. Section \ref{4} then contains our explicit expressions for both the bulk dual of a fluid with $n$ anomalous currents and the bulk dual of the CME; these are the main results of our paper. Section \ref{5} has a discussion of our results and we end with an appendix reviewing the Weyl covariant formalism introduced in \cite{Loganayagam} which we make extensive use of.

\section{Perturbative Construction of Event Horizon} \label{2}
The aim of this section and the subsequent one is to review the methodology involved in determining the location of the event horizon and in constructing the dual entropy current, as first introduced in \cite{Harvey}. These calculations are insensitive to many of the details specific to each individual fluid dynamical bulk metric, so we will work with as much generality as possible for now, but will introduce the specific bulk solution to a fluid with anomalous currents in section \ref{4}. 
\subsection{Generic Fluid Dynamical Metric}
We must first fix a gauge. We choose the following gauge which is consistent with previous work \cite{B2,Ashok}.\footnote{We should add that other papers on the fluid/gravity correspondence \cite{Erdmenger,Banerjee,Harvey} have used other gauges, but at first order in boundary derivatives this is not really an issue.}
\begin{equation} \label{gauge}
g_{rr}=0, \quad g_{r\mu} =- u_\mu.
\end{equation}
Here, $r$ refers to the radial coordinate while Greek indices label boundary coordinates. The vector field $u_\mu$ is the fluid dynamical velocity.  Also, note that the bulk metric $g_{AB}$ admits an expansion in boundary spacetime derivatives; 
\begin{equation}
g_{AB}=g_{AB}^{(0)} + g_{AB}^{(1)} + g_{AB}^{(2)} + \cdots,
\end{equation}
where $g^{(m)}_{AB}$ represents the term of order $m$ in boundary derivatives. This reflects the \textit{long wavelength} nature of the boundary fluid dynamics - the fluid dynamical parameters must be slowly varying with respect to the equilibriation scale and thus the fluid dynamical stress tensor is naturally expressed as an expansion in derivatives of the velocity $u_\mu$ and temperature $T$ fields to a specified order.

Consistent with our gauge choice, we can parametrise our bulk metric in the following manner:
\begin{equation} \label{InitialBulk}
ds^2=-2 u_\mu dx^\mu dr  +  \chi_{\mu\nu} (r,x) dx^\mu dx^\nu.
\end{equation}
As expected, the function $\chi_{\mu\nu}$ is expressed in a series organised in increasing order of boundary derivatives, as follows:
\begin{equation}
\chi_{\mu\nu}=\chi_{\mu\nu}^{(0)} + \chi_{\mu\nu}^{(1)} + \chi_{\mu\nu}^{(2)} + \cdots.
\end{equation}
It is worth pausing here to mention the geometric interpretation of our bulk metric parametrisation. In this gauge, lines of constant $x^\mu$ are ingoing null geodesics with $r$ an affine parameter along them. This congruence of null geodesics provides a natural way of mapping quantities defined on the horizon to the boundary - we will make use of this in section \ref{3}.

In the following subsection, we proceed to determine the location of the event horizon for general fluid dynamical metrics of the form (\ref{InitialBulk}).

\subsection{Local Event Horizon}
In this paper, we are interested in fluid flows on flat boundary metrics. Given the dissipative nature of the Navier-Stokes equations, it is reasonable to assume that such flows will eventually settle down to a uniform configuration described by constant parameters $u_0$ and $T_0$. The holographic dual of such a system is simply a uniform black brane. Determining the location of the event horizon at late times is therefore straightforward - it must reduce to the event horizon of a uniform black brane, $r_U$, which we assume to be known. Using this, we can now determine the event horizon of a general fluid dynamical metric: it is the unique null hypersurface that reduces to $r_U$ at late times.

This asymptotic condition can easily be incorporated into our concept of a boundary derivative expansion. If we assume that the event horizon is defined by $S_\mathcal{H}(r,x)=0$ where
\begin{equation} \label{horizon}
\begin{aligned}
S_\mathcal{H}(r,x)&=r-r_H(x),\\
r_H(x)&=r_0(x) +r_1(x)+r_2(x) + \cdots
\end{aligned}
\end{equation}
with $r_m$ representing the term of order $m$ in boundary derivatives, then choosing $r_0(x)=r_U(T(x))$ precisely accounts for our late time condition. When our fluid globally thermalises and settles down into uniform flow, all higher order terms $r_m$ with $m\ge 1$ will vanish and the location of the event horizon will be given by $r=r_U$, as required. 

The higher order terms, $r_m$ for $m\ge 1$, can then be determined by requiring that the event horizon, defined by $S_\mathcal{H}(r,x)=0$, be a null hypersurface. The normal to this hypersurface is, by definition, $\xi^A=g^{AB}\partial_B S_\mathcal{H}(r,x)$. The requirement that it be null is simply given by:
\begin{equation} \label{null}
g^{AB}\partial_A S_\mathcal{H} \partial_B S_\mathcal{H}=0,
\end{equation}
where the inverse metric $g^{AB}$ for our general fluid dynamical metric (\ref{InitialBulk}) is as follows:
\begin{equation}
g^{rr} = \frac{1}{-\,  u_\mu \, u_\nu \, \chi^{\mu\nu}} \ , \quad
g^{r\alpha} = \frac{ \, \chi^{\alpha\beta}\,u_{\beta}}{-  \, u_\mu \, u_\nu \, \chi^{\mu\nu}} \ , \quad
g^{\alpha\beta} = \frac{ \, u_ \gamma \, u_ \delta \, \left(
\chi^{\alpha\beta} \,\chi^{\gamma\delta} - \chi^{\alpha \gamma} \,\chi^{\beta \delta}  \right)}{-  \, u_\mu \, u_\nu \, \chi^{\mu\nu}} \ ,
\label{invmet}
\end{equation}	
with $\chi^{\mu\nu}$ specified by $\chi_{\mu\nu} \, \chi^{\nu\rho} = \delta_\mu^{\ \rho}$. In terms of these components, equation (\ref{null}) can be expressed as:
\begin{equation}
0 =  \frac{1}{- \,  u_\mu \, u_\nu \, \chi^{\mu\nu}} 
\left( 1 + 2  \,  \, \chi^{\alpha\beta}\,u_{\beta} \, \partial_\alpha S_H 
-   \, \left(
\chi^{\alpha\beta} \,\chi^{\gamma\delta} - \chi^{\alpha \gamma} \,\chi^{\beta \delta}  \right) 
\, u_ \gamma \, u_ \delta \, \partial_ \alpha S_H \, \partial_ \beta S_H.
\right)
\end{equation}	
Solving this equation at each order in the boundary derivative expansion then gives us the location of our event horizon.

At first sight, this local procedure may seem counter-intuitive. The event horizon is in general a nonlocal concept and determining its location usually requires a complete understanding of  the future evolution of the spacetime. Yet, here, we have obtained the location of the event horizon at a given point, $r(x)$, in terms of the derivatives of the fluid dynamical parameters at the corresponding boundary point - a manifestly local construction. This is not as surprising as it may seem though. And the reason for this lies with the fact that fluid dynamical spacetimes are well approximated tube-wise by uniform black branes; this is a reflection of the field theory's local equilibrium. Within the regime where this remains a good approximation - a region of size\footnote{The characteristic scale set by local equilibriation is of order $1/T$.} $1/T$ around $x^\mu$ -  we know that the geodesic trajectories must be similar to those of a uniform black brane. It is therefore possible to determine the location of the event horizon, $r(x)$, using just the fluid dynamical data in a region of size $1/T$ around the corresponding boundary point. Based on this explanation, we should still expect a limited degree of nonlocality which does not seem to be present in our local perturbative procedure. However, upon summing the terms $r_m$ in (\ref{horizon}) to all orders in boundary derivatives, we expect our expression for the location of the event horizon to become nonlocal, relying on data within a patch of size $1/T$. The resulting nonlocal expression which is still fairly ateleological, is an intriguing property of fluid dynamical metrics, signifying that the event horizon acts like a replica of the boundary fluid, locally mirroring its behaviour \cite{Harvey}.

\section{Local Entropy Current} \label{3}
In this section, we explain how to holographically construct an entropy current for the dual fluid. Roughly speaking, we utilise the area form on the event horizon to define an entropy current, and then we map this to the boundary, thus obtaining an entropy current for the dual fluid.

\subsection{Coordinates Specific to Event Horizon}
It is straightforward to construct the area form on the event horizon if we make a judicious choice of coordinates. Here, we will do exactly this. In the following subsection, we will then reexpress this area form in the global coordinates described by (\ref{InitialBulk}).

Recall that the event horizon is a codimension one null hypersurface in the $(d+1)$-dimensional bulk spacetime. Its normal vector therefore lies within its tangent space. Further, the integral curves generated by the normal vector exactly cover the event horizon. We choose as our coordinates $(\lambda,\alpha^a)$ such that $\lambda$ is a future directed parameter along these integral curves and that $\alpha^a$ is orthogonal to the other tangent vectors and is itself null, the metric restricted to the event horizon takes the following simple form:
\begin{equation} \label{hormetf}
ds^2= \, \mathfrak{g}_{ab}\,  d\alpha^a \, d\alpha^b.
\end{equation}
We can easily define an area form on the event horizon (appropriately normalised to coincide with the entropy):
\begin{equation} \label{enttf1} 
a=\frac{1}{4\, G_{d+1}} \, \sqrt{\mathfrak{g}} \, d\alpha^1\wedge d\alpha^2 \wedge \ldots \wedge d\alpha^{d-1},
\end{equation}
where $\mathfrak{g}$ represents the determinant of the $(d-1)\times (d-1)$ metric  $\mathfrak{g}_{ab}$.
Equation (\ref{enttf1}) gives us an expression for the area form in the coordinates $(\lambda,\alpha^a)$ which are specific to the event horizon; we now proceed to rewrite this in terms of more global coordinates.

\subsection{Global Coordinates}
We aim to relate the coordinates $(\lambda,\alpha^a)$ introduced in the previous subsection to the global coordinates $(r,x^\mu)$ for fluid dynamical metrics defined by (\ref{InitialBulk}). 

Consider the spacelike hypersurface specified by $x^0=\lambda$. Such a surface will intersect each integral curve generated by the normal vector to the event horizon exactly once. We can therefore take $\lambda$ to be a parameter along these curves. We further choose the value of $x^a$, for $a=1,\cdots,d-1$, at $\lambda=0$ to label each integral curve - we refer to this value as $\alpha^a$. Note that these coordinates $(\lambda, \alpha^a)$ are an explicit example of the coordinates described in the previous subsection. It is straightforward to express these coordinates in term of $x^\mu$ in the vicinity of $\lambda=0$:
\begin{equation}\label{coordchange} 
\begin{split}
x^a& = \alpha^a+ \frac{n^a}{n^\lambda}\, \lambda + \frac{\lambda^2}{2 \,n^\lambda}\, 
 n^\mu\,\partial_\mu \left( \frac{n^a}{n^\lambda}\right) + {\cal O}(\lambda^3) \cdots, \\
d x^a&=d \alpha^a + \lambda \,d \alpha^k \, \partial_k \left( 
\frac{n^a}{n^\lambda} \right) +  d \lambda \,\left( \frac{n^a}{n^\lambda} + \frac{\lambda}{n^\lambda}  \,
n^\mu\,\partial_\mu\left( \frac{n^a}{n^\lambda} \right)\right) + {\cal O}(\lambda^2) ,
\end{split}
\end{equation}
where $(n^\lambda,n^a)$ represents the normal vector to the event horizon. We now use these relations to express the area form (\ref{enttf1}) in terms of the coordinates $x^\mu$. 

We first need to express the metric on the event horizon $\mathfrak{g}_{ab}$ in terms of the global metric $g_{AB}$ of (\ref{InitialBulk}). Let $h_{\mu \nu} \,dx^\mu \, dx^\nu  = g_{AB}\, dx^A\, dx^B|_\CH$ denote the metric restricted to the event horizon  in the $x^\mu$ coordinates. We therefore have that:
\begin{equation}\label{metexp}
\begin{split}
ds_\CH^2 &= h_{\mu\nu}(x)\, dx^\mu \,dx^\nu \equiv \mathfrak{g}_{ab} \,d\alpha^a 
\,d\alpha^b\\
&=
h_{i j} \! \left(\lambda,\alpha^i+\frac{n^i}{n^\lambda}\lambda\right) \, 
\left( d \alpha^i +\lambda \, d \alpha^k \, \partial_k  \left( \frac{n^i}{n^\lambda}
\right) \right)
\left( d \alpha^j +\lambda\, d \alpha^k \,\partial_k  \left( \frac{n^j
}{n^\lambda}
\right) \right) +{\cal O}(\lambda^2) 
\end{split}
\end{equation}
where $h_{ij}(\lambda,x)$ represents $h_{\mu\nu}$ restricted to a constant-$\lambda$ hypersurface.
The determinant of $\mathfrak{g}_{ab}$ is thus given by:
\begin{equation}\label{detmet}
\sqrt{\mathfrak{g}}= \sqrt{h}|_{\lambda=0} + \frac{\lambda}{n^\lambda}  \, 
\left( n^i \,\partial_i \sqrt{h} + \sqrt{h}\, n^\lambda \, \partial_i \frac{n^i}{n^\lambda} 
\right) +{\cal O}(\lambda^2) \ , 
\end{equation}
where $h|_{\lambda=0}$ is the determinant of $h_{ij}$ restricted to $\lambda=0$. 

It is now easy to obtain a global expression for the area $(d-1)$-form:
\begin{equation}\label{enttf2}
a=\frac{\sqrt{\mathfrak{g}}}{4\,G_{d+1}} \, d\alpha^1\wedge d\alpha^2 \ldots \wedge d \alpha^{d-1}.
\end{equation}
We simply have to substitute for $\sqrt{\mathfrak{g}}$ and $d\alpha^a$ using (\ref{coordchange}) and (\ref{detmet}). The resulting expression is, however, unnecessarily complicated because of the expansions in $\lambda$. If we choose to restrict this expression to $\lambda=0$, we obtain the following simpler formula:
\begin{equation}\label{enttf3}
a=\frac{\sqrt{h}}{4\,G_{d+1}} \,  \left(  dx^1\wedge dx^2 \ldots \wedge d x^{d-1}
-\sum_{i=1}^{d-1}\, \frac{n^i}{n^\lambda} \, d \lambda \wedge dx^1 \wedge \ldots \wedge dx^{i-1} \wedge dx^{i+1} \wedge \ldots  \wedge dx^{d-1} \right) |_{\lambda=0}
\end{equation}
Observe that the above expression can be rewritten as a current as follows:
\begin{equation}\label{enttf4} 
a=\frac{\epsilon_{\mu_1 \mu_2 ... \mu_{d}}}{(d-1)!} \; 
J_S^{\mu_1} \, dx^{\mu_2} \wedge ...\wedge dx^{\mu_{d}}|_{\lambda=0}.
\end{equation}
where $J_S^\mu$ is given by  
\begin{equation}\label{encur}
J^\mu_S= \frac{\sqrt{h}}{4\,G^{(d+1)}_N}\,  \frac{n^\mu}{n^\lambda}.
\end{equation}
Although we have derived this expression specifically for $\lambda=0$, note that our splitting of $x^\mu$ into $(\lambda,x^a)$ and our choice of origin for $\lambda$ were completely arbitrary. The resulting expression for the entropy current\footnote{This expression is actually independent of the splitting of $x^\mu$ into $(\lambda,x^a)$ eventhough this is not explicitly manifest - see  section 3.3 of \cite{Harvey}.} (\ref{encur}) is therefore valid anywhere on the horizon.

And finally, we would like to transport this current from the event horizon (where it is currently defined) to the boundary. The most natural way of doing this is to map the value of the current at $(r(x),x^\mu)$ on the horizon to the boundary point $(r=\infty,x^\mu)$ by following the null geodesics given by $x^\mu=$ constant. We now have an entropy current for the dual fluid. The area increase theorem of black holes then guarantees the positivity of the divergence of this current.

\section{Explicit Results to First Order in Boundary Derivatives} \label{4}
Here, we present the explicit results of our calculations; these are the main results of this paper. In the first subsection, we give the expressions for the bulk metric, (outer) event horizon, and entropy current for the fluid/gravity model with $n$ anomalous currents. These are already novel results which further generalise existing literature. In the second subsection, we then specify to the fluid/gravity model of the chiral magnetic effect - these results are of particular interest from a condensed matter perspective.

\subsection{Fluid/Gravity Model with $n$ Anomalous Currents}

We aim to construct bulk gravitational solutions dual to a fluid with $n$ anomalous currents in the presence of background electromagnetic fields. This can be done by obtaining long wavelength solutions to the bulk dynamics described by the $U(1)^n$ Einstein-Maxwell-Chern-Simons action\footnote{Repeated lower case Latin indices imply summation, for example:
\begin{equation}
q^a q^a \equiv \Sigma_{a=1}^n q^a q^a.
\end{equation} 
.}
\begin{equation}
S=\frac{1}{16\pi G_5} \int d^5x \sqrt{-g}\left[R-12-F_{MN}^a F^{aMN}+\frac{S^{abc}}{6\sqrt{-g}}\epsilon^{PKLMN}A_P^a F_{KL}^b F_{MN}^c\right].
\end{equation}
The triangles anomalies for the currents are encoded in the Chern-Simons parameter via the relation:
\begin{equation}\label{anomaly}
 C^{abc}=-\frac{S^{abc}}{4\pi G_5}
\end{equation}
as found in \cite{Surowka}. 

The equations of motion which result from this action are given by:
\begin{equation} \label{eqofmot}
\begin{aligned}
  G_{MN} - 6 g_{MN} + 2 \left( F^a_{MR} F^{aR}{}_N - \frac{1}{4} g_{MN}
    F^a_{SR} F^{aSR} \right)  &= 0,
  \\
  \nabla_M F^{aMP} = -\frac{S^{abc}}{8\sqrt{-g}} \varepsilon^{PMNKL}
  F^b_{MN} F^c_{KL} &.
\end{aligned}
\end{equation}
The bulk fluid duals can then be computed using the method pioneered in \cite{B1} (see also \cite{R2,R3,B2}). We now only briefly outline this method, referring the reader to the original references for further details. First, note that the equations (\ref{eqofmot}) admit the following solution describing a uniform charged black brane corresponding to a field theory state at constant temperature and chemical potentials:
\begin{equation}
\begin{aligned}
  ds^2 &= - 2u_\mu dx^\mu dr - r^2f(r)u_\mu u_\nu dx^\mu dx^\mu + r^2 P_{\mu\nu} dx^\mu dx^\nu , \\
  A^a  &= \frac{\sqrt{3} q^a}{2r^2} u_\mu dx^\mu \,, 
\end{aligned}
\end{equation}
where
\begin{equation}
\begin{aligned}
  f(r)&= 1 - \frac{m}{r^4} +\frac{q^a q^a}{r^6}\,,\\
P_{\mu\nu}&=\eta_{\mu\nu}+u_\mu u_\nu.
\end{aligned}
\end{equation}
The function $f(r)$ has two positive, real solutions which describe the outer and inner horizon of the black brane. The position of the (outer) event horizon is at $r=r_+$ with
\begin{equation}
r_+ =\frac{\pi T}{2}\left(1+\sqrt{1+\frac{8}{3\pi^2}\frac{\mu^a\mu^a}{T^2}}\right).
\end{equation}
There is also an inner horizon at $r=r_-$: 
\begin{equation}
r_-^2=\frac{1}{2} r_+^2\left(-1+\sqrt{9-\frac{8}{\frac{1}{2}\left(1+\sqrt{1+\frac{8}{3\pi^2}\frac{\mu^a\mu^a}{T^2}}\right)}}\right)\,.
\end{equation}
Here, the chemical potentials $\mu^a$ and temperature T can be expressed in terms of bulk quantities as:
\begin{equation} \label{chem}
\begin{aligned}
T&=\frac{r_+}{2\pi}\left(2-\frac{q^a q^a}{r_+^6}\right) ,\\
\mu^a&=\frac{\sqrt{3}q^a}{2r_+^2}.
\end{aligned}
\end{equation}
To obtain a bulk hydrodynamic dual representing a locally equilibriated boundary configuration, we proceed by promoting the constant mass $m$, charges $q^a$, and velocity field $u^\mu$ to slowly-varying fluid dynamical parameters; we further introduce background gauge fields $\mathfrak{A}^a_\mu$ to model external electromagnetic fields $B^{a\mu}=\frac{1}{2}\epsilon^{\mu\nu\alpha\beta}u_\nu F_{\mathfrak{A}\alpha\beta}^a$ and $E^{a\mu}=F^{a\mu\nu}_\mathfrak{A} u_\nu$ (where $F^{\mu\nu}_\mathfrak{A}=\partial^\mu\mathfrak{A}^{a\nu}-\partial^\nu\mathfrak{A}^{a\mu}$). We therefore take the following expressions as our zeroth order ansatz:
\begin{equation}
\begin{aligned}
  ds^2 &= - 2u_\mu (x) dx^\mu dr - r^2f(r,m(x),q^a(x))u_\mu (x) u_\nu (x) dx^\mu dx^\mu + r^2 P_{\mu\nu} (x) dx^\mu dx^\nu , \\
  A^a  &= \frac{\sqrt{3} q^a (x)}{2r^2} u_\mu (x) dx^\mu + \mathfrak{A}_\mu (x) dx^\mu\,, 
\end{aligned}
\end{equation}
and perturbatively solve the equations (\ref{eqofmot}) to a specified order in boundary derivatives.

To first order, we obtain the following expression for the bulk metric:
\begin{equation}\label{bulk}
\begin{aligned}
  ds^2 &= - 2u_\mu dx^\mu \left(dr+r\mathcal{A}_\nu dx^\nu\right) - r^2f(r)u_\mu u_\nu dx^\mu dx^\mu + r^2 P_{\mu\nu} dx^\mu dx^\nu , \\
&-\frac{2\sqrt{3}S^{abc}q^a q^b q^c}{8mr^4}u_{(\mu}\omega_{\nu)} dx^\mu dx^\nu
-\frac{12r^2}{r_+^7}u_{(\mu} P_{\nu)}^\lambda q^a  \mathcal{D}_\lambda q^a F_q (r) dx^\mu dx^\nu \\
&-2F_B(r) S^{abc}q^a q^b u_{(\mu} B^c_{\nu )} dx^\mu dx^\nu \\
&-2F_E (r) q^a u_{(\mu}E^a_{\nu)} dx^\mu dx^\nu +\frac{2r^2}{r_+}F_\sigma(r) \sigma_{\mu\nu} dx^\mu dx^\nu
\end{aligned}
\end{equation}
where
\begin{equation}
\begin{aligned}
F_q(r)&\equiv\frac{1}{3}f(r)\int_r^\infty dr^\prime \frac{r_+^8}{f(r^\prime)^2}\left(\frac{r_+}{r^{\prime 8}}-\frac{3}{4r^{\prime 7}}\left(1+\frac{r_+^4}{m}\right)\right),\\
F_\sigma (r)&\equiv\int_\frac{r}{r_+}^\infty dr^\prime \frac{r^\prime\left(r^{\prime 2}+r^\prime+1\right)}{\left(r^\prime +1\right)\left(r^{\prime 4}+r^{\prime 2} - \frac{m}{r_+^4} +1\right)},\\
F_B(r)&=3r^2 f(r)\int_\infty^r dr^\prime \frac{1}{r^{\prime 5} f(r^\prime)^2}\left(\frac{1}{2r^{\prime 4}} - \frac{r_+^6+\frac{q^a q^a}{4}}{r^{\prime 2}r^4_+ m}\right),\\
F_E(r)&= -r^2f(r)\int_r^\infty dr^\prime \left(\frac{3\sqrt{3} q^a q^a}{r_+ m r^{\prime 3} }-\frac{4\sqrt{3}(r^\prime-r_+)}{r^{\prime 3}}\right)\int_{r^\prime}^\infty dr^{\prime\prime} \frac{1}{r^{\prime\prime 5} f(r^{\prime\prime})^2}\\
&+r^2f(r)\left(\int_r^\infty dr^\prime \frac{1}{r^{\prime 5} f(r^\prime)^2}\right)\left(\int_r^\infty dr^\prime\left(\frac{3\sqrt{3} q^a q^a}{r_+ m r^{\prime 3} }-\frac{4\sqrt{3}(r^\prime-r_+)}{r^{\prime 3}}\right)\right).
\end{aligned}
\end{equation}
The Weyl covariant derivative $\mathcal{D}_\mu$, the Weyl connection $\mathcal{A}_\mu$, and the Weyl covariant (pseudo)tensors $\sigma_{\mu\nu}$, $\omega^\mu$ are defined in Appendix \ref{A}. We now pause to elaborate on the significance of these results within the context of existing literature. The results above describe the bulk dual of a fluid with \textit{multiple anomalous currents} in the presence of \textit{background electromagnetic fields}. While similar situations have been studied in the literature \cite{Kirsch, Erdmenger,Banerjee,Hur,Megias}, their calculations were for more specific configurations. Our results, which are also explicitly Weyl covariant, extend existing work on the fluid/gravity correspondence to greater generality.

With this bulk metric in hand, we can determine the location of the outer horizon, $r_H=r_+ + r_1 + \cdots$, following the procedure laid out in section \ref{2}. However, we find that the first order term $r_1$ vanishes. In hindsight, this is unsurprising and follows from Weyl covariance \cite{Harvey}. Bulk fluid dynamical spacetimes are invariant under boundary Weyl transformations of the hydrodynamic variables - this is a reflection of the conformal nature of the boundary fluid dynamics. This boundary Weyl symmetry constrains the form of the terms allowed in the bulk metric. Consequently, the location of the horizon must be given as a sum of Weyl covariant scalars; and since there are no Weyl covariant scalars at first order in boundary derivatives, the outer horizon receives no first order correction.
 
And finally, applying the method discussed in section \ref{3}, we compute the entropy current to first order in boundary derivatives:
\begin{equation}\label{ec}
\begin{aligned}
J_S^\mu&=\frac{1}{4G_5}\left(r_+^3 u^\mu + \frac{3q^a (q^b q^b +2r_+^6)}{4m(2r_+^6-q^b q^b)}P^{\mu\nu} \mathcal{D}_\nu q^a- S^{abc}q^a q^b q^c \frac{\sqrt{3}}{8 m r_+^3}\omega^\mu \right. \\
&- \left. S^{abc}q^a q^b  \frac{3}{8mr_+}B^{c\mu} + \frac{\sqrt{3}\pi(2r_+^6-q^b q^b)}{2m^2}q^aE^{a \mu}\right).
\end{aligned}
\end{equation}
To make contact with existing literature, we rewrite our expression for the entropy current in the following form:
\begin{equation}
J_S^\mu=su^\mu - \frac{\mu^a}{T}\nu^{a\mu} + D \omega^\mu +D^a_B B^{a\mu},
\end{equation}
where
\begin{equation}
\nu^{a\mu}=-\sigma^{ab} T P^{\mu\nu}\partial_\nu\left(\frac{\mu^b}{T}\right)+\sigma^{ab} E^{b\mu}+\xi^a\omega^\mu+\xi_B^{ab} B^{b\mu}.
\end{equation}
The entropy density $s$, electrical conductivity $\sigma^{ab}$, and anomalous transport coefficients $\xi^a$ and $\xi^{ab}_B$ are given by:
\begin{equation}
\begin{aligned}
s&=\frac{r_+^3}{4G_5}\\
\sigma^{ab}&=\frac{\pi T^2r_+^7}{4G_5m^2}\delta^{ab}\\
\xi^a&= -\frac{3S^{abc}q^b q^c}{16\pi G_5}\\
\xi^{ab}_B &=-\frac{\sqrt{3}\left(3 r_+^4 +m\right) S^{abc} q^c}{32\pi G_5 m r_+^2}
\end{aligned}
\end{equation}
The expressions for $s$ and $\sigma_{ab}$ can be obtained by comparison with (\ref{ec})\footnote{We make use of the relations (\ref{chem}).}; the anomalous transport coefficients $\xi^a$ and $\xi^{ab}_B$ were holographically computed in \cite{Surowka}. We assume that these parameters are given and use our holographic computation of the entropy current (\ref{ec}) to determine the coefficients $D$ and $D_B^a$. We find that:
\begin{equation}
\begin{aligned}
D&=\frac{1}{3T}C^{abc}\mu^a\mu^b\mu^c,\\
D_B^a&=\frac{1}{2T}C^{abc}\mu^b\mu^c.
\end{aligned}
\end{equation}
This is in perfect agreement with the results of \cite{Surowka}; here, the authors computed the coefficients $D_B^a$ and $D$ using thermodynamic arguments. Our results thus provide holographic confirmation of their calculation.

\subsection{Fluid/Gravity Model of the Chiral Magnetic Effect}
In this subsection, we specify to the fluid/gravity model of the chiral magnetic effect. As explained in the introduction, we require just two currents, an axial current and a vector current, with associated gauge fields $A^A$ and $A^V$, and chemical potentials $\mu_5$ and $\mu$. An external magnetic field is needed as well. To model this, it is sufficient to have just an external vector gauge field $\mathfrak{A}^{V\mu}$ with a nonzero magnetic component $B^\mu$; we set the corresponding electric field $E^\mu$ to zero. The anomaly coefficient $C^{abc}$ is determined by $C-$parity \cite{Kirsch2}:
\begin{equation}
\begin{aligned}
C^{121}&=C^{211}=C^{112}=0\,, \\
C^{222}&=0 \,, \\
C^{111}&\neq 0 \,, \\
C^{122}&=C^{221}=C^{212}\neq 0 \,. 
\end{aligned}
\end{equation}
We choose $C^{111}=C^{122} \equiv C/3$; this then fixes the Chern-Simons parameter $S^{abc}$ via relation (\ref{anomaly}).

Observe that, with the restrictions imposed thus far, the two currents are both anomalous. This is not consistent with the hydrodynamic model of the CME which requires just the axial current to be anomalous, accounting for the imbalance in chirality caused by the axial anomaly. The vector current should remain conserved. Thish issue can easily be addressed by the addition of Bardeen currents \cite{Kirsch2} which can be incorporated by including the Bardeen counterterm in the action; the Bardeen currents will restore conservation of the vector current. However, here, we are primarily interested in the form of the coefficients which appear in the expression for the entropy current. These coefficients depend only on the temperature $T$ and chemical potentials $\mu$ and $\mu_5$, and are unchanged by the inclusion of Bardeen currents. We therefore neglect this subtlety in our calculation.

The bulk metric is similar to (\ref{bulk}) with just the inclusion of the modifications described above, so we do not write it here explicitly. And, similarly, there is no first order correction to the location of the outer horizon. For the entropy current, we find the following expressions for the coefficients $D$ and $D_B^a$:
\begin{equation}
\begin{aligned}
D&=\frac{C}{T}\mu^2\mu_5,\\
D_B&=\frac{C}{T}\mu\mu_5.
\end{aligned}
\end{equation}
 And this is completely consistent with the results of the purely hydrodynamic analysis of \cite{Sad}.

\section{Discussion} \label{5}
In this paper, we have constructed long wavelength asymptotically locally AdS$_5$ bulk spacetimes with slowly varying gauge fields which are solutions of the $U(1)^n$ Einstein-Maxwell-Chern-Simons system. These bulk spacetimes are dual to $(3+1)-$dimensional fluid flows with $n$ anomalous currents in the presence of background electromagnetic fields. Using the methodology introduced in \cite{Harvey}, we then computed the entropy current for the boundary fluid flow to first order in boundary spacetime derivatives. These results further generalise existing work on the fluid/gravity correspondence. In particular, holographic computations of the entropy current have previously only been done for a fluid with a single $U(1)$ current (possessing a $U(1)^3$ anomaly) without any background electromagnetic fields present. Finally, we restricted our results to the fluid/gravity model of the chiral magnetic effect, and calculated the corresponding entropy current. Our calculations are all consistent with those done using conventional thermodynamic/hydrodynamic methods.

A very natural generalisation of our work would be to extend our bulk metric (\ref{bulk}) to second order in boundary derivatives. Then, using standard holographic techniques, the corresponding transport coefficients can easily be obtained. It would certainly be of interest to understand the physical significance of these second order coefficients. 

The logical next step would be to extend the expression for the entropy current to second order as well; however, this is a much more intricate problem. From a purely hydrodynamic perspective, the requirement that the divergence of the entropy current be non-negative fixes the first order contribution to the current. This is not true in general at second order and an ambiguity remains. This has been shown explicitly in \cite{BB}. Here, the author found that the entropy current (with non-negative divergence) at second order for an uncharged conformal fluid has a four-parameter ambiguity. We should also mention here that these results can also be reproduced by demanding consistency with a partition function \cite{Shiraz} instead of using the constraints arising from non-negativity of the divergence of the entropy current. And in fact, this alternative method may be more computationally tractable. It would be very useful to extend the analyses of \cite{BB,Shiraz} to a charged fluid with anomalous currents, as studied in this paper.

If we now consider this issue holographically, the ambiguity in the entropy current at second order should be reflected in the bulk construction as well. Two sources of ambiguity have already been determined in previous literature. First, there exists freedom in how we choose to map the gravitational entropy current from the horizon to the boundary \cite{Harvey}. And second, it is possible to construct entropy currents (with non-negative divergence) on horizons other than the event horizon - this leads to a further source of ambiguity. In \cite{Spalinski,Spalinski2}, the authors constructed an entropy current on a Weyl-invariant apparent horizon and demonstrated that it also satisfies all relevant hydrodynamic constraints. A possible extension of our work would be to utilise their method to construct this alternative entropy current. It would be interesting to see if the ambiguity in the entropy current computed holographically precisely matches the ambiguity one would find using the hydrodynamic analyses of \cite{BB,Shiraz}. We will revisit these issues in future work.

\acknowledgments
It is a pleasure to thank Harvey Reall for comments on a draft of this paper.

\appendix

\section{Weyl Covariant Formalism} \label{A}
In this appendix, we first briefly outline the Weyl covariant formalism introduced in \cite{Loganayagam}. We then define several Weyl covariant tensors that were used in this paper and we also classify all Weyl covariant tensors of interest (up to rank two) at first order in boundary derivatives.

 We begin by explaining the significance of Weyl covariance for bulk fluid duals. For the bulk metrics constructed in the fluid/gravity correspondence, the AdS asymptotics imply that the boundary fluid possesses conformal symmetry. This invariance under boundary Weyl transformations constrains the form of the bulk metric  - the terms must transform in an appropriate Weyl covariant manner. It is thus useful to utilise a formalism where Weyl covariance is explicitly manifest. The main obstacle towards achieving this is that ordinary covariant derivatives of Weyl covariant tensors are not usually Weyl covariant. The key idea in \cite{Loganayagam} was to introduce a `Weyl covariant derivative'; we now elaborate further on this.

A Weyl covariant tensor is a quantity that transforms homogeneously under a Weyl transformation. More specifically, a tensor of weight $w$ transforms in the following manner:
\begin{equation}
\mathcal{Q}^{\mu\cdots}_{\nu\cdots}=e^{-w\chi(x)}\tilde{\mathcal{Q}}^{\mu\cdots}_{\nu\cdots}
\end{equation}
under a Weyl rescaling, $g_{\mu\nu}=e^{2\chi(x)}\tilde{g}_{\mu\nu}$. The Weyl covariant derivative introduced in \cite{Loganayagam} acts on a Weyl covariant tensor of weight $w$ as follows: 
\begin{equation} \label{Weylderivative}
\begin{aligned}
\mathcal{D}_\lambda \mathcal{Q}^{\mu\cdots}_{\nu\cdots} &\equiv \nabla_\lambda \mathcal{Q}^{\mu\cdots}_{\nu\cdots} + w \mathcal{A}_\lambda \mathcal{Q}^{\mu\cdots}_{\nu\cdots} \\
&\quad + \left[g_{\lambda\alpha}\mathcal{A}^\mu-\delta_\lambda^\mu\mathcal{A}_\alpha - \delta_\alpha^\mu \mathcal{A}_\lambda \right] \mathcal{Q}^{\alpha\cdots}_{\nu\cdots} +\cdots \\
&\quad - \left[g_{\lambda\nu}\mathcal{A}^\alpha-\delta_\lambda^\alpha\mathcal{A}_\nu - \delta_\nu^\alpha \mathcal{A}_\lambda \right] \mathcal{Q}^{\mu\cdots}_{\alpha\cdots} -\cdots.
\end{aligned}
\end{equation}
The Weyl connection, $\mathcal{A}_\mu$, is constructed from the fluid velocity field, $u_\mu$, as given below:
\begin{equation}
\mathcal{A}_\mu \equiv u^\lambda \nabla_\lambda u_\mu - \frac{\nabla_\lambda u^\lambda}{d-1} u_\mu .
\end{equation}
It is not difficult to show that the Weyl covariant derivative of a tensor of weight $w$ is itself a Weyl covariant tensor of weight $w$. This formalism thus allows us to maintain explicit Weyl covariance.

We will now define several Weyl covariant tensors that were introduced in section \ref{4}: the shear tensor $\sigma_{\mu\nu}$ and the vorticity (pseudo)tensors, $\omega_{\mu\nu}$ and $\omega^\mu$,
\begin{equation}
\begin{aligned}
\sigma_{\mu\nu} &\equiv \mathcal{D}_{( \mu} u_{\nu ) }, \\
\omega_{\mu\nu} &\equiv \mathcal{D}_{[\mu} u_{\nu ] } ,\\
\omega^\mu &\equiv\epsilon^{\mu\nu\alpha\beta}u_\nu\omega_{\alpha\beta}.
\end{aligned}
\end{equation}
And finally, we list all independent\footnote{By independent, we mean that the fluid dynamical equations of motion impose no relations between these tensors.} Weyl covariant scalars, transverse (pseudo)vectors, symmetric traceless transverse tensors at first order in boundary derivatives; such terms are of importance for the bulk metric, the location of the horizon, and the computation of the entropy current. It turns out that there are no Weyl covariant scalars, three Weyl pseudovectors $\omega^\mu$, $B^\mu$, and $E^\mu$, and just one symmetric transverse traceless tensor, $\sigma_{\mu\nu}$.

\end{document}